# Complex Design Networks: Structure and Dynamics†

Dan Braha

New England Complex Systems Institute, Cambridge, MA 02138, USA
University of Massachusetts, Dartmouth, MA 02747, USA
Email: braha@necsi.edu

Why was the $6 billion FAA air traffic control project scrapped? How could the 1977 New York City blackout occur? Why do large scale engineering systems or technology projects fail? How do engineering changes and errors propagate, and how is that related to epidemics and earthquakes? In this paper we demonstrate how the rapidly expanding science of complex design networks could provide answers to these intriguing questions. We review key concepts, focusing on non-trivial topological features that often occur in real-world large-scale product design and development networks; and the remarkable interplay between these structural features and the dynamics of design rework and errors, network robustness and resilience, and design leverage via effective resource allocation. We anticipate that the empirical and theoretical insights gained by modeling real-world large-scale product design and development systems as self-organizing complex networks will turn out to be the standard framework of a genuine science of design.

## 1. Introduction: the road to complex networks in engineering design

Large-scale product design and development is often a distributed process, which involves an intricate set of interconnected tasks carried out in an iterative manner by hundreds of designers (see Fig. 1), and is fundamental to the creation of complex man-made systems [1—16]. This complex network of interactions and coupling is at the heart of large-scale project failures as well as of large-scale engineering and software system failures (see Table 1).

Many within the design community contributed to the understanding and representation of the complex interdependencies characterizing product design and development [17—30]. While this effort led to very important insights and effective applications, the approach largely involved computational techniques applied to relatively small design networks, which hindered the discovery of universal non-trivial topological patterns across design networks from multiple domains, their striking similarity with other social, biological, and technological networks; and, most importantly, the direct interplay between these universal structural patterns and the functionality of design networks. Thus, a new interdisciplinary scientific approach that is concerned with empirical as well as theoretical studies of real-world large-scale engineering design networks was needed in order to understand the relationship between network architectures (topologies) and network dynamics [11—15]—a critical step towards effective management of complex design products and projects, and the prevention of engineering failures.

---

†This paper is based on keynote lectures delivered on October 12, 2009 at the 11[th] International DSM Conference in Greenville, South Carolina, and on August 6, 2013 at the 39[th] Design Automation Conference in Portland, Oregon. A previous version of this paper appeared as Braha D. 2016. The Complexity of Design Networks: Structure and Dynamics. In Cash P, Stankovic T, and Štorga M. (Eds.), Experimental Design Research: Approaches, Perspectives, Applications. Springer.

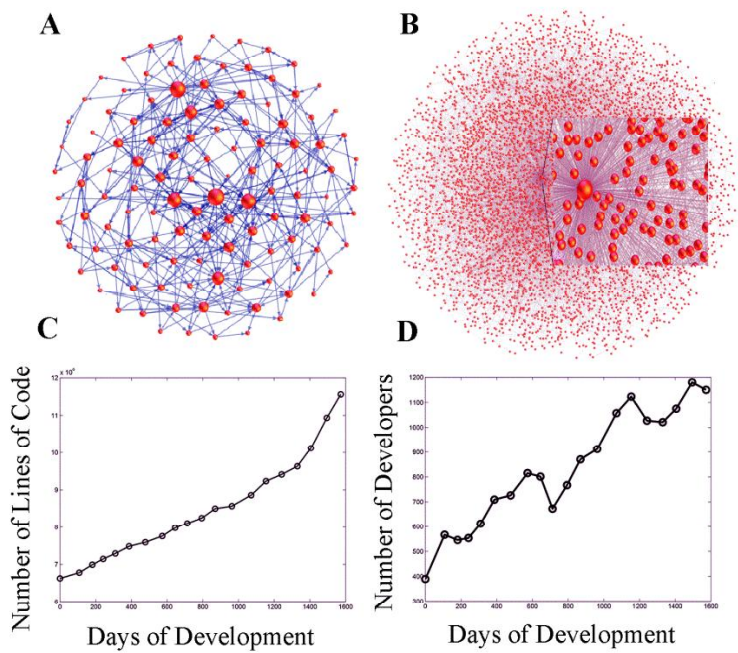

**Figure 1. Complex Product Design and Development Networks.** (**A**) Network of information flows between main tasks of a vehicle large-scale design [11—15]. This task network consists of 417 directed information flows between 120 development tasks. Each task is assigned to one or more actors (design teams, engineers, or scientists) who are responsible for it. Here, the information links are directed – each task consumes information from others and generates information to others. (**B**) Open source software system [14]. The software system network was generated from the call graphs of the Linux operating system kernel (version 2.4.19). A call graph is a directed graph that represents calling relationship among subroutines. This software network consists of 11,460 directed information flows between 5,420 subroutines. In both networks, the degree of a node (i.e., the number of nodes adjacent to a node) is represented by the size of the node. The design networks were visualized using Gephi 0.8.2. (**C**) Linux kernel development: size of source code over product development time. The Linux kernel keeps growing in size over time as more hardware is supported and new features are added [16]. Software size is measured as lines of code. Each data point represents a different Linux version (beginning with version 2.6.11, and ending with version 2.6.30; see [16]). (**D**) Linux kernel development: number of developers over product development time. The number of Linux developers (and likely interaction among them) shows an increasing trend over the different Linux kernel versions.

This paper is a brief introduction to the emergent complex networks theory approach to the understanding of the structure and dynamics of real-world complex product design and development networks [11—15, 31]. In this framework, a complex network is a graph (network) with non-trivial structural properties that are ubiquitous in a wide variety of real-world large-scale networks. For product design and development networks, the nodes of the network could represent people, tasks, subroutines or logic gates, which communicate via links representing engineering change orders, parameters, specifications, or signals.

The field of complex design networks is largely inspired by the empirical study of very large real-world product design and development networks including, for example, forward logic chips with 23,843 logic gates and 33,661 signal links, open source software systems with 5,420 subroutines and 11,460 calling relationships among subroutines, or a product development process with 889 tasks and 8,178 information flows. The empirical study of such product design and development networks has led to many surprising non-trivial structural (architectural) properties—similar to those found in other biological, social, and technological networks—which indicates that real-world design networks are distinctive, and are neither purely regular nor purely random. Such features (defined later in the paper) include sparseness of connections, short average path lengths between any two nodes, high-levels of clustering and modularity, 'fat-tailed' degree centrality distributions and the existence of very highly connected nodes (critical nodes or 'hubs'), asymmetric information flows, disassortativity among nodes, community structure, and hierarchical network organization.

The raison d'être, though, of complex network studies is to understand the relationship between structure and dynamics. The dynamics of product design and development networks can be understood to be due to processes propagating through the network of connections, including the propagation of changes, errors, and defects. Remarkably, the universal structural properties found in complex design networks provide key information about the characteristics of error and defect propagation, both whether and how rapidly it occurs. Moreover, these architectural properties have implications for the functional utility of product design and development systems, including their robustness (error tolerance) to failures and unintended design changes, and their sensitivity to planned resource allocation.

This complex networks approach to the study of design is gaining increasing attention within the design community, who confirmed the theory and extended its scope in important ways—for example, from the measurement of centrality and the identification of critical nodes [32—36]; to the role of critical nodes in system's quality [33], project's risk propagation [34, 36], and prioritization of resource allocation [35]; to the relationship between network assortativity, cycles, or loops on product quality [37]; to the detection of network modular structures and communities [38—40]; and understanding network's robustness against failures and attacks [41—42].

**Table 1. Large Scale Product Design and Project Failures.**

| System | Failure |
|---|---|
| Columbia Space Shuttle, 2003 | Damage to thermal protection tiles, leading to left wing structural failure |
| The New York blackout of 1977 | Multiple lightning strikes at Buchanan South substation, tripping two circuit breakers. |
| Mars Climate Orbiters, 1999 | Mixture of pounds and kilograms, leading to the failure of the software controlling the orbiter's thrusters |
| Pentium II and Pentium Pro FPU bug, 1994 | Incomplete entries in a look-up-table used by the floating-point division circuitry, processor can return incorrect decimal results |
| Gulf of Mexico oil spill, 2010 | Sea-floor oil gusher following the explosion and sinking of the Deepwater Horizon oil rig |
| US Federal Aviation Administration Advanced Automation System, 1982-1994 | Project was abandoned in 1994 with an estimated cost of $6B |
| London Stock Exchange Taurus Paperless Stock Trading System, 1990-1993 | Project was abandoned in 1993 with an estimated cost of $600M |
| US Air Force Advanced Logistics System, 1968-1975 | Project was abandoned in 1975 with an estimated cost of $250M scrapped |

## 2. The Universality of Complex Networks

Networks have become a standard model for a wealth of complex systems, from physics to social sciences to biology [43—44]. A large body of work has investigated topological properties [43] including changes due to node removal [45—47]. The main objective, though, of complex network studies is to understand the relationship between structure and dynamics [48]—from disease spreading and social influence [49—52] to search [53] and time dependent networks [54, 55]. Complex networks theory has also contributed to engineering environments, where new theoretical approaches and useful insights from application to real data have been obtained [8]. Most importantly, these structural patterns and dynamical properties were found to be universal—that is, the same or very similar in a wide variety of complex systems (see Figure 2). Basic definitions and notations of networks pertinent to this paper are described in the Appendix (The reader is recommended to read it first).

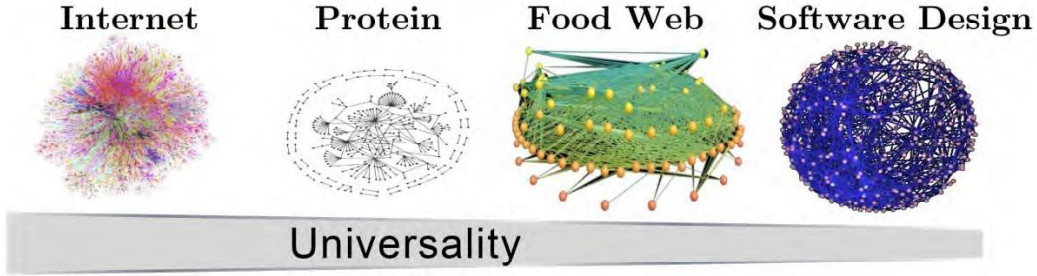

**Figure 2. The Universality of Complex Networks.** Network Patterns are found to be the same in a wide variety of technological, biological, and social systems. This paper demonstrates that engineering design networks can be put in the same class as complex networks in other domains.

Of particular interest are scale free networks where the degree (i.e., the number of nodes adjacent to a node) is distributed according to a power law or a long right tail distribution (see Appendix). Such networks have characteristic structural features like 'hubs', highly connected nodes [43], features which cause them to exhibit super-robustness against failures [45—47] on the one hand, and super-vulnerability to deliberate attacks and epidemic spreading [49—52] on the other. Here, we find that the framework of complex networks, mainly applied to natural, social, and biological systems, can be usefully applied and extended to understand the relationship between the structure and dynamics of large scale design networks.

Regular networks, where all the degrees of all the nodes are equal (such as circles, grids, and fully connected graphs) have been traditionally employed in modeling physical systems of atoms [56]. On the other hand, many 'real-world' social, biological and technological networks appear more random than regular [43—44]. With the scarcity of large-scale empirical data on one hand and the lack of computing power on the other hand scientists have been led to model real-world networks as completely random graphs using the probabilistic graph models of Erdős and Rényi [57].

In their seminal paper on random graphs, Erdős and Rényi [57] considered a model where $N$ nodes are randomly connected with probability $p$. In this model, the average degree of the nodes in the network is $\langle k \rangle \cong pN$, and a Poisson distribution approximates the distribution of the nodal degree. In a Poisson random network, the probability of nodes with at least $k$ edges decays rapidly for large values of $k$. Consequently, a typical Poisson random network is rather homogenous, where most of the nodal degrees are concentrated around the mean. In particular, the average distance between any pair of nodes (the 'characteristic path length,' see Appendix) scales with the number of nodes as $d_{\text{random}} \sim \ln(N)/\ln(\langle k \rangle)$. This feature of having a relatively short path between any two nodes, despite the often large graph size, is known as the small-world effect. In a Poisson random graph, the clustering coefficient (see Appendix) is $C_{\text{random}} = p \cong \langle k \rangle/N$. Thus, while the average distance between any pair of nodes grows only logarithmically with $N$, the Poisson random graph is poorly clustered.

Regular networks and random graphs serve as useful models for complex systems; yet, many real networks are neither completely ordered nor completely random. It has been found that social, technological, and biological networks are much more highly clustered than a random graph with the same number of nodes and edges (i.e., $C_{\text{real}} \gg C_{\text{random}}$), while the characteristic path length $d_{\text{real}}$ is close to the theoretically minimum distance obtained for a random graph with the same average connectivity [43—44]. Small-World Networks are a class of graphs that are highly clustered like regular graphs ($C_{\text{real}} \gg C_{\text{random}}$), but with a small characteristic path length like a random graph ($d_{\text{real}} \approx d_{\text{random}}$). Many real-world complex systems have been shown to be small-world networks, including power-line grids, neuronal networks, social networks, the World-Wide Web, the Internet, food webs, and chemical-reaction networks.

Another important characteristic of real-world networks is related to their node degree distribution (see Appendix). Unlike the bell-shaped Poisson distribution of random graphs, the degree distribution of many real-world networks has been documented to follow a power-law,

$$p(k) \sim k^{-\gamma} \qquad (1)$$

where $p(k)$ is the probability that a node has $k$ edges (or neighbors). Networks with power-law distributions are often referred to as scale-free networks [43—44]. A power-law distribution is an example of an uneven node degree distribution, which is characterized by a long right tail—some nodes are very highly connected ('hubs'), while most have small degrees. These heavy-tailed distributions are characterized by 'wild' variability and right-skewness of the connectivity distributions. The term 'wild' variability means that the second moment $\langle k^2 \rangle$ (equivalently the variance) of the degree distributions is extremely large (and sometimes diverges) relative to the average degree of the nodes in the network. This is in contrast to the fast decaying tail of a Poisson distribution, which results in a small second moment or variance. Power-law distributions of both the in-degree and out-degree of a node have also been observed in a variety of directed real-world networks [43—44] including the World-Wide Web, metabolic networks, networks of citations of scientific papers, and telephone call graphs. Although scale-free networks are prevalent, the power-law distribution is not universal. Empirical work shows that the node degree distribution of a variety of real networks often has a scale-free regime with an exponential cutoff, i.e. $p(k) \sim k^{-\gamma} e^{-\left(\frac{k}{k^*}\right)}$ where the parameter $k^*$ is the cutoff of the degree distribution [58]. The existence of a cutoff has been attributed to physical costs of adding links or limited capacity of a vertex [58]. In some networks, the power-law regime is not even present and the node degree distribution is characterized by a distribution with a fast decaying tail. Moreover, studies of the dynamics of link utilization in complex networks offer a radical alternative to the static-based view of complex networks [54—55]. In such time-dependent networks, there is hardly any continuity in degree centrality of nodes over time (that is, hubs rarely stay hubs for any length of time); even though cross-sectional snapshots are scale-free networks.

## 3. Complex Design Networks: Structural Properties

The goal of the present section is to describe several of the key statistical properties of large-scale design networks. We show that large-scale design networks, although of a different nature, have general properties that are shared by other social, technological, and biological networks.

First, it is found that complex design networks are highly sparse, that is, they have only a small fraction of the possible number of links (i.e., have low density; see Appendix). The low sparseness of design networks (see Table 2) implies that the functionality of these networks (e.g., effective information flow between designers) is not related to the sheer number of information links in the system but to the way those information flows are patterned in the network. We will substantiate this observation more formally in Section 4.

Moreover, complex design networks are 'small-world' networks; that is, despite being primarily locally connected and modular, such design networks exhibit short average path lengths between any two nodes. This is shown in Table 3 where we compare the clustering coefficients and characteristic path lengths of several large-scale real design networks with the corresponding characteristics computed from a random ensemble of random graphs with the same number of nodes and links. We see that the clustering coefficients of the real networks are much higher than the clustering coefficients of the corresponding random graphs ($C_{\text{real}} \gg C_{\text{random}}$), but with similar characteristic path lengths ($d_{\text{real}} \approx d_{\text{random}}$). A high clustering coefficient is consistent with a modular organization; that is, the organization of the system (project or product design) in clusters that contain most, if not all, of the interactions internally, while minimizing the interactions or links between separate clusters. However, while 'modularity' is intuitively perceived as inversely related to the rate of information transfer throughout the network; here we see that 'small-world' design networks have the capacity of fast information transfer, which results in immediate response to signals propagated from other components of the product design, or rework created by other tasks in a product development network (see Section 4).

In Section 2, we considered two typical network topologies: Poisson random networks and scale-free networks. Statistical analysis of the data reveals an asymmetric pattern of node degree distributions related to the information flowing into and out of nodes (product design components or product development tasks). More specifically, both the degree distributions of incoming and outgoing information flows show a power-law regime with a decaying tail (see Figure 3). However, the degree distributions related to the incoming information flows seem to exhibit a faster decaying tail (much like a Poisson distribution), whereas the degree distributions related to the outgoing information flows seem to be highly heterogeneous (much like a power-law distribution, see Figure 3). The noticeable asymmetry between the distributions of incoming and outgoing information flows shown by large-scale design networks suggests that the incoming capacities of nodes (e.g., the ability to integrate and process information) are much more limited than their counterpart outgoing capacities. The power-law behavior of the incoming and outgoing distributions suggests that nodes play distinct roles in processing information flows. More specifically, it implies that the dynamics of directed design networks is dominated by a few *highly connected hubs*, which either consume and/or generate a lot of information through network links. These are the *'information bottlenecks'* (hubs) of the design network.

The functional significance of the power-law behavior of the incoming and outgoing distributions is intimately linked to two important characteristics of design networks: 'ultra-robustness' and 'ultra-leverage.' Ultra-robustness is the ability of a design network to be resilient and error tolerant when unexpected and negative design changes occur over time; while ultra-leverage is the ability to influence the performance of the system (measured, for example, in terms of defects or product development time) by taking advantage of the 'wild' variability and right-skewness properties of the incoming and outgoing connectivity distributions. More specifically, a remarkable improvement in the performance of engineering design systems can be achieved by focusing engineering and management efforts on central 'information-consuming' and 'information-generating' nodes. We further elaborate on these issues in Section 5.

**Table 2. Density of Real-World Design Networks.** Complex design networks are highly sparse.

|  | Network | Type | # Nodes | # Links | Density |
|---|---|---|---|---|---|
| Open-Source Software | Linux-kernel [14] | Directed | 5,420 | 11,460 | $3.9 \cdot 10^{-4}$ |
|  | MySQL [14] | Directed | 1,501 | 4,245 | $19 \cdot 10^{-4}$ |
| Forward Logic Chip | s38417 electronic circuit [14] | Directed | 23,843 | 33,661 | $5.9 \cdot 10^{-5}$ |
|  | s38584 electronic circuit [14] | Directed | 20,717 | 34,204 | $7.9 \cdot 10^{-5}$ |
| Product Development | Vehicle [12—14] | Directed | 120 | 417 | $2.9 \cdot 10^{-2}$ |
|  | Operating software [12—14] | Directed | 466 | 1,245 | $5.7 \cdot 10^{-3}$ |
|  | Pharma facility [12—14] | Directed | 582 | 4,123 | $1.2 \cdot 10^{-2}$ |
|  | 16 Story hospital facility [12—14] | Directed | 889 | 8,178 | $10^{-2}$ |
| Technological | Internet [59] | Undirected | 21,823 | 44,933 | $1.8 \cdot 10^{-4}$ |
|  | Power grid [43, 44] | Undirected | 4,941 | 6,594 | $2.7 \cdot 10^{-4}$ |

**Table 3. The 'Small-World' Property of Complex Design Networks**. While the random graphs are not modular (low clustering coefficients), design networks exhibit the 'small-world' property of high degree of modularity (high clustering coefficients) and fast information transfer among nodes (short average path lengths between any two nodes).

|  | **Network** | $d_{\text{real}}$ | $d_{\text{rand}}$ | $C_{\text{real}}$ | $C_{\text{rand}}$ |
|---|---|---|---|---|---|
| **Open-Source Software** | Linux-kernel | 4.66 | 5.87 | 0.14 | 0.001 |
|  | MySQL | 5.47 | 4.20 | 0.21 | 0.004 |
| **Forward Logic Chip** | s38417 electronic circuit | 20.66 | 23.48 | 0.016 | 0.0001 |
|  | s38584 electronic circuit | 13.39 | 17.32 | 0.012 | 0.00003 |
| **Product Development** | Vehicle | 2.88 | 2.70 | 0.21 | 0.07 |
|  | Operating software | 3.70 | 3.45 | 0.33 | 0.02 |
|  | Pharma facility | 2.63 | 2.77 | 0.45 | 0.02 |
|  | 16 Story hospital facility | 3.12 | 2.58 | 0.27 | 0.02 |
| **Technological** | Internet | 4.11 | 3.95 | 0.37 | 0.001 |
|  | Power grid | 18.7 | 12.4 | 0.08 | 0.005 |

We conclude this section by introducing two concepts that are important in understanding the dynamics of complex design networks: assortativity and dissortativity. Assortativity (or assortative mixing) refers to the tendency of nodes in a network to connect to other nodes with similar properties. Here, we focus on assortativity in terms of a node's degree. That is, a network is assortative if it is likely that nodes with similar degrees (low or high) connect to each other. That is, assortative mixing by degree is observed in networks that exhibit positive correlations between the degrees of neighboring nodes. On the other hand, a network is disassortative if it is likely that high degree nodes connect to low degree nodes. Disassortative mixing by degree is observed in networks that exhibit negative correlations in their degree connectivity patterns.

The concept of assortativity (or dissortativity) in the context of directed networks (typical for design networks) can be extended by considering several mixing patterns in the network (see Figure 4). As shown in the second column of Figure 4, assortative (or disassortative) mixing can also be observed at the level of individual nodes. In the assortative case, we check if nodes that have low in-degrees (respectively high in-degrees) tend to also have low out-degrees (respectively high out-degrees). ; that is, whether the network exhibits a positive correlation between the in-degree and out-degree of nodes. The assortative (or disassortative) patterns of the network provide full information on the structure of the network [14]. If the network is uncorrelated (neither assortative nor disassortative), the only relevant information for the structure of the network is the node degree distribution $p(k)$, or in case of directed networks the in- and out-degree distributions $p_{\text{in}}(k)$ and $p_{\text{out}}(k)$.

The presence and the extent of mixing patterns in a network have a profound effect on the topological properties of the network as it affects the detailed wiring of links among nodes. It is also closely related to the dynamics of error and change propagation in large-scale design networks as discussed in Section 4. For example, assortative mixing (positive correlations) leads to complex structural properties including cycles, loops, and the emergence of a single connected component (referred to as the Giant Component, see Appendix) that contains most of the nodes in the network (and thus many loops). These structural features tend to amplify the propagation of design changes and errors through the design network. It is thus expected that efficient design networks show negative (or no) correlations in their degree connectivity patterns. This is, indeed, empirically observed as shown in Figure 4. In Section 4 we provide a theoretical explanation of this empirical fact. We conclude the section by noting that, in the context of design networks, all of the above observations (e.g., the existence of hubs or dissortativity) are not limited to degree centrality measures but also hold when other node centrality indices of Social Networks Analysis (SNA) are applied, including closeness centrality, betweenness centrality, or eigenvector centrality [12].

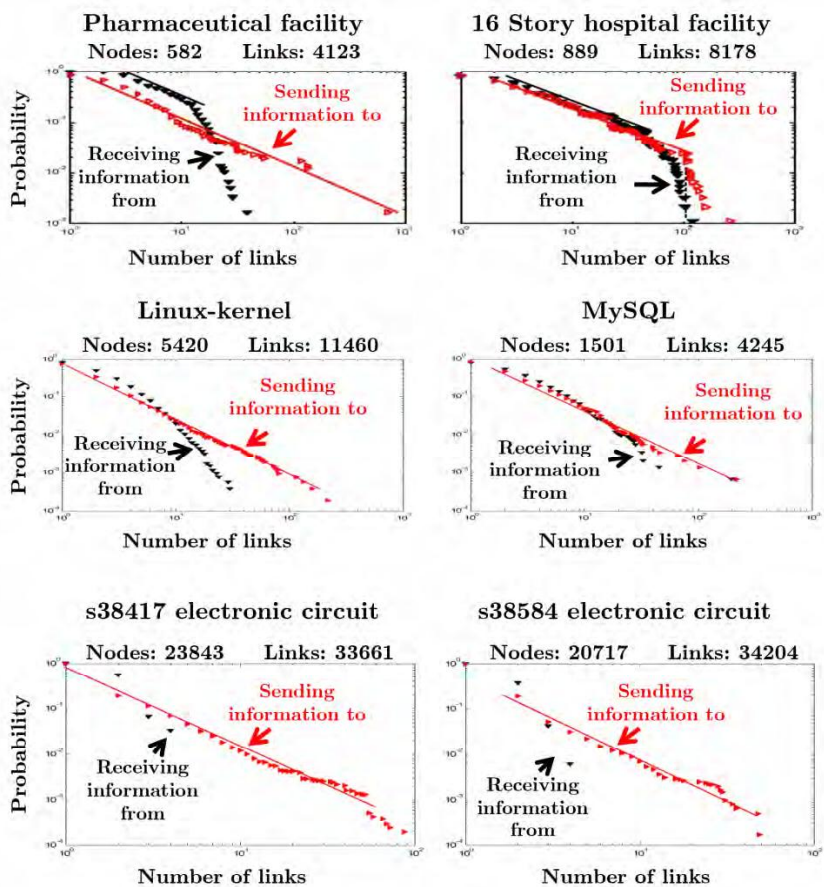

**Figure 3. Degree Distributions of Complex Design Networks**. Here we plot on a logarithmic scale the inverse cumulative degree distributions (i.e. Prob(Degree ≥ $k$)). While both the incoming ('receiving information from') and outgoing ('sending information to') connections of nodes show a power-law regime

(straight line on a logarithmic scale) with a decaying tail, the incoming link distributions have sharp cutoffs that are substantially lower than those of the outgoing link distributions [12—14].

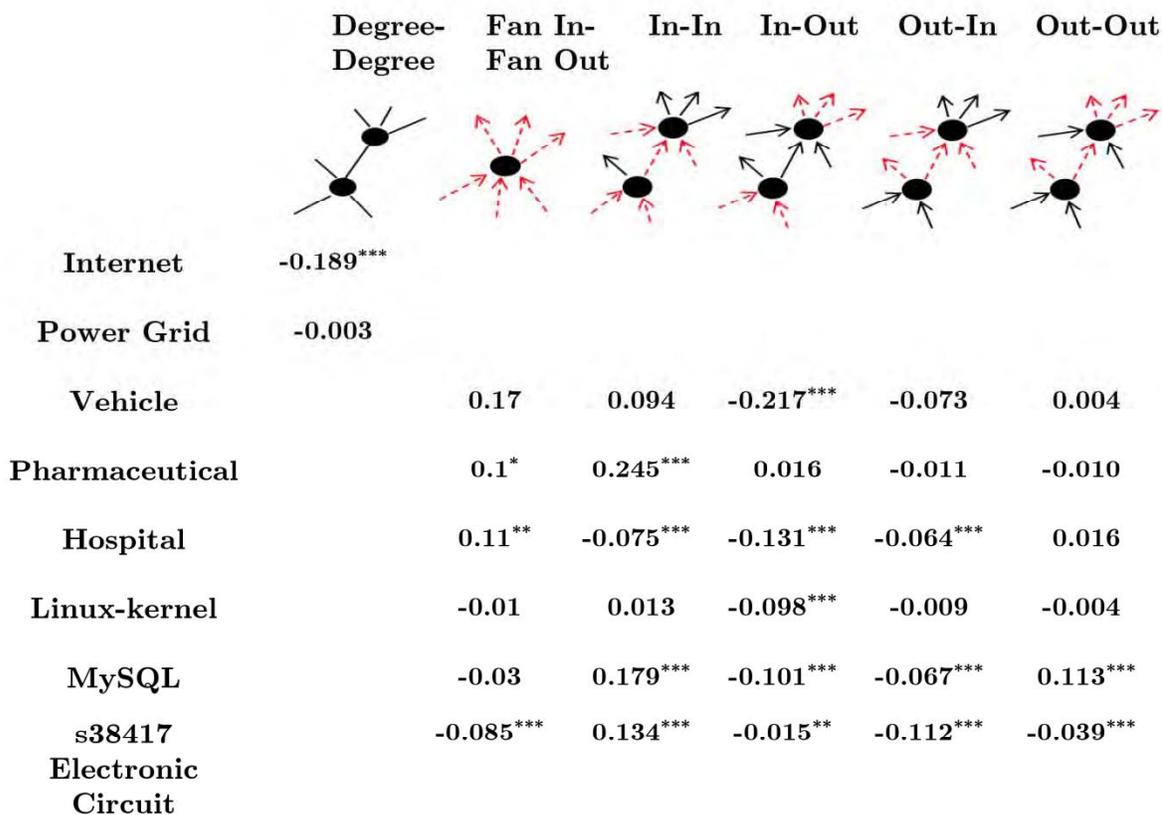

**Figure 4. Degree Correlations in Design Networks.** The reported numbers are the Pearson correlation coefficients for various mixing patterns in the network [12—14]. Notice that for directed networks, several different mixing patterns can exist depending on the directionality of links. We also denote whether the reported correlations are statistically different from zero at the *5%, **1%, or ***0.1% level. Overall, the results provide support for the hypothesis that complex design networks exhibit negative (or zero) correlations in their degree connectivity patterns, a finding explained in terms of network dynamics (see Section 4).

## 4. Error and Change Propagation in Complex Design Networks

In this section, we present a model for the dynamics of errors, rework, or change propagation in complex design networks. Here we outline basic results; a detailed account of the dynamic network model is given by [14]. Consider a scenario of designing a complex engineering system (e.g. airplane, car, software), which involves a large number of development teams. As shown in Figure 5A, we consider a network representing development tasks carried out by teams who work to resolve various open design problems. The network includes $N$ nodes taking only the values 0 (colored red in Figure 5) or 1 (colored blue in Figure 5), representing 'open' or 'resolved' state of a particular task, respectively. At each time step, a node is selected at random. If the node is in a 'resolved' state (Figure 5B, top), its state can be modified depending on the 'open' nodes connected to it through incoming links. These 'open' nodes send out change order information that might lead to the reopening of the 'resolved' task. More specifically, each incoming 'open' task causes the 'resolved' task to reopen its state with probability $\beta$ (the 'Coupling Coefficient'). If the node is in an 'open' state (Figure 5B, bottom), its state can be modified depending on two conditions: (1) it is not affected by any of its incoming 'open' tasks (each with probability $1 - \beta$), and (2) it becomes 'resolved' with probability $\delta$ (the 'Recovery Coefficient'). The latter condition captures the idea that development teams can resolve the open problems autonomously, regardless of the states of incoming nodes. Though not an essential assumption, in order to gain insight into the model, we assume that $\beta_i = \beta$ and $\delta_i = \delta$ for all nodes in the network—considered as typical average values.

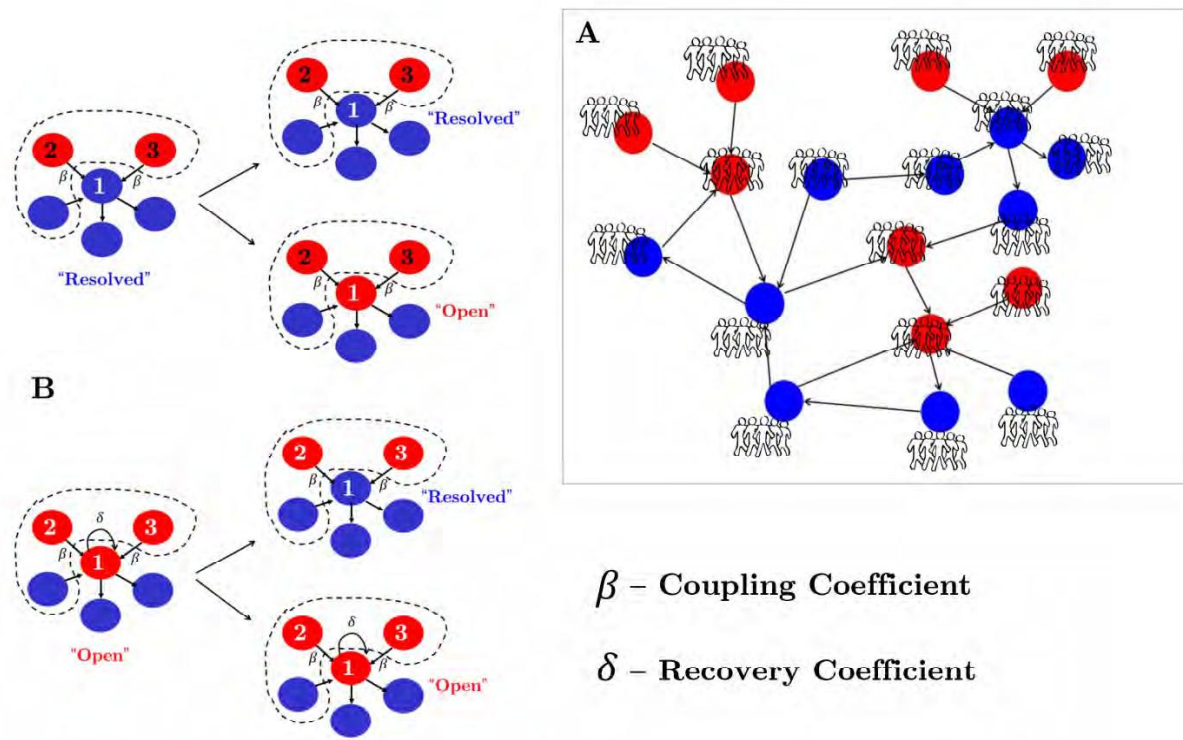

**Figure 5. A dynamic network model of error propagation in complex design networks. (A)** The design network consists of nodes representing development tasks carried out by teams who work to resolve various open design problems. The teams interact with one another via communication links. In the diagram, blue and red nodes represent 'resolved' and 'open' tasks, respectively. **(B)** The stochastic rules that govern the dynamics of the system. The model involves two parameters, which measure the coupling strength between neighboring nodes, $\beta$; and the rate by which development teams resolve open problems autonomously, $\delta$.

As the project unfolds, open tasks are resolved autonomously and later in time might be reopened in light of the influence of 'open issues' that are propagated via to neighboring unresolved tasks in the network; thus generating additional rework and revision. This process continues until either all tasks become 'resolved' or until the network settles into an equilibrium state of *non-zero* fraction of 'open' tasks. The latter outcome is an undesirable result from a project management perspective.

To illustrate this dynamical behavior, we show in Figure 6A two typical simulation runs of the dynamic network model. The underlying network in this case is the real-world pharmaceutical product development network, which includes 582 nodes (tasks) and 4213 links (see Table 2). The bottom graph (Figure 6A, circle marker type) shows the time evolution of the percentage of open nodes, leading to a converging network where there are no open tasks in the network. Increasing the coupling (i.e. increasing the coefficient $\beta$) between neighboring nodes in the network leads to a different qualitative behavior as shown in the top graph of Figure 6A (square marker type). In this case, the project spirals out of control with open problems remaining indefinitely in the network. These two different types of behaviors will be explained by the theory presented below. It is instructive to compare the simulation results to the dynamics of open problems observed in real product development projects. In Figure 6B, we show the dynamics of open problems surveyed in a family of vehicle programs (interior and exterior subsystem design) at a large automotive company [20]. The similarity in dynamical behavior between the model and real-world data is appealing.

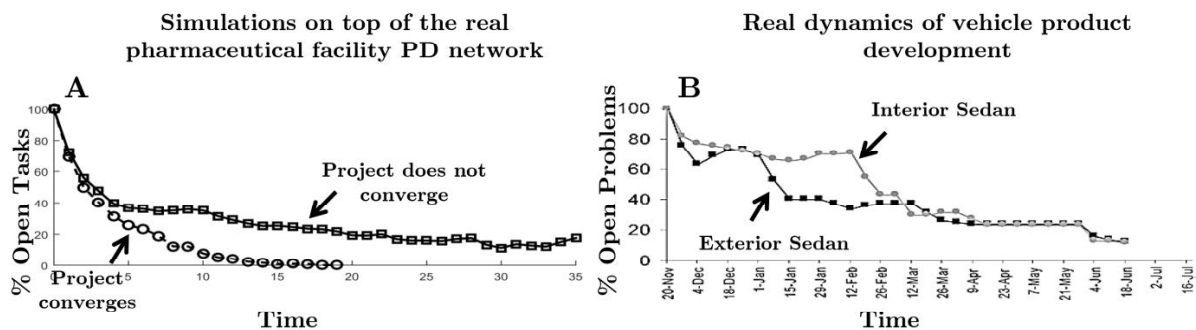

**Figure 6. The percentage of open problems in simulated and real product development systems**. (**A**) A typical simulation run of the dynamic network model on a real world pharmaceutical product development network (see Table 2) with 582 nodes (tasks). The average number of in-coming arcs connected to a node is 7.08. The bottom graph (circle marker type) shows the time evolution of the percentage of open nodes when the coupling and recovery coefficients are $\beta = 0.065$ and $\delta = 0.5$, respectively. In this case, the simulation run converges to the fully-resolved state where there are no open tasks in the project. The top graph (square marker type) shows the time evolution when the coupling and recovery coefficients are $\beta = 0.09$ and $\delta = 0.5$, respectively. In this case, the increase in coupling between neighboring nodes (from $\beta = 0.065$ to $\beta = 0.09$) leads to a project that spirals out of control with open problems remaining indefinitely in the project network. Both of these outcomes can be predicted by our theory. (**B**) The dynamics of open problems observed in a family of vehicle programs based on real-world data collected at a large automotive company [20].

The dynamic network model can represent a variety of real engineering design systems. An example is the propagation of failures in complex engineered systems such as power grids, communication systems, computer networks, or mechanical structures. In this case, the state of a node (e.g., a power substation) could represent its maximum working capacity. When a node in the network fails, it transfers its load to neighboring nodes in the network. Those neighboring nodes are then become overloaded and transfer their load to other nodes, triggering cascading failures throughout the system.

Next we address the following key question: given the coupling strength between neighboring nodes $\beta$, the recovery coefficient of self-directed problem solving $\delta$, the initial number of 'open' tasks, and the underlying network structure, how will the fraction of 'open' tasks develop with time? Most importantly, will the system converge to the globally resolved state, where the fraction of 'open' tasks becomes zero? or perhaps, over the long run, will there always be a fraction of 'open' nodes and open problems present in the network? Remarkably, it is shown that the structural properties (discussed in Section 3) of the underlying network provide key information about the characteristics of error, rework, or defect dynamics on top of this network. Here, we represent network structure by considering the various correlations in the degree connectivity patterns of the network as shown in Figure 4.

We begin our analysis by investigating the effect of an Erdős-Rényi random network (See Section 2) on the dynamics of error propagation. For a random network, the degree correlations corresponding to the mixing patterns shown in Figure 4 are absent. Despite of the fact that real design networks are different from random networks (as discussed in Section 3), the analysis of this special case will be useful for understanding the dynamics of error propagation on general networks. The main result is summarized in Figure 7A. The long-term behavior of the system is determined by whether $\delta \geq \beta \langle k \rangle$ or $\delta < \beta \langle k \rangle$, regardless of the initial number of 'open' tasks. In the former case, the project converges to the globally resolved state, where the fraction of 'open' tasks becomes zero; otherwise, the project spirals out of control with persistent 'open' tasks in the network. We thus have a threshold phenomenon.

This threshold behavior is further illustrated by the phase diagram shown in Figure 7B. The phase diagram shows the conditions at which distinct phases can occur at equilibrium. Here, the x-axis shows the coupling strength between neighboring nodes $\beta$, and the y-axis shows the recovery coefficient of self-directed problem solving $\delta$. The two phases in the diagram are separated by the line $\delta = \beta \langle k \rangle$. So, imagine a project corresponding to point 1 in Figure 7B. The conditions specified by point 1 imply that the project will converge to the globally resolved state. Increasing the coupling between tasks (point 2 in Figure 7B) will still lead to project convergence, though the time to complete the project may be longer. Increasing the average connectivity $\langle k \rangle$ of the network will increase the slope of the line that separates the two phases in the diagram (Figure 7C). In this case, starting at point 1 (corresponding to a convergent project) and increasing the coupling between tasks may lead to a project that spirals out of control (point 2 in Figure 7C). Thus, for a given $\beta$ and $\delta$, adding more links (and thus more complexity) to a project network can impede the project's convergence.

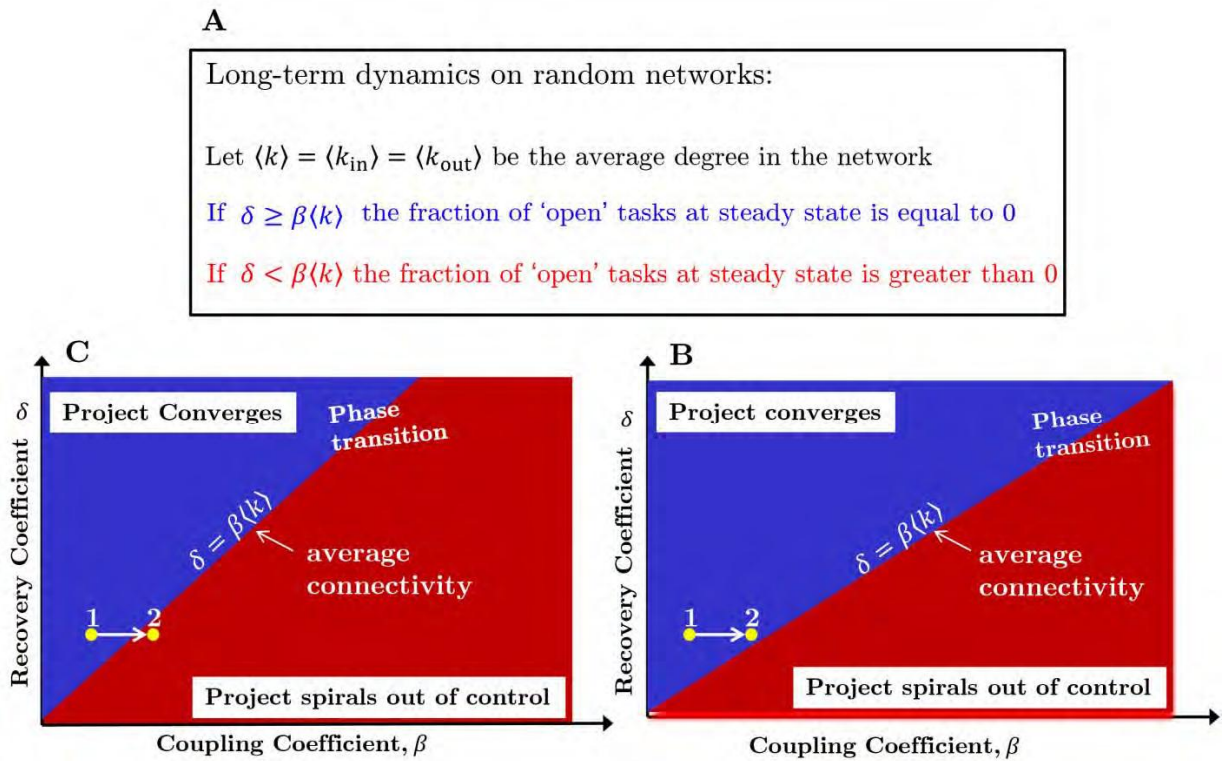

**Figure 7. The dynamics of error propagation on Erdős-Rényi random networks.** **(A)** The long-term dynamics of the network is determined by a threshold that depends on the coupling coefficient, recovery coefficient, and the average connectivity in the network. **(B)** and **(C)** demonstrate the threshold behavior by using phase diagrams, which show the conditions at which the two distinct phases can occur at equilibrium.

We next analyze the dynamics of error propagation for correlated networks (for which the random network is a special case). In general, it can be shown [14] that the dynamics is determined by the degree correlations corresponding to the various mixing patterns in the directed design network (top row in Figure 4). In this paper, however, we focus on a special class of correlated networks where the only relevant information is related to the correlation between the in- and out-degrees of individual nodes (the fan-in/fan-out mixing pattern in the second column of Figure 4). This approximation is fully justified for many design networks where the observed correlations between neighboring nodes are very small (this is indeed the case as shown in Figure 4).

The main result is summarized in Figure 8A. The long-term behavior of the system is now determined by whether $\delta \geq \beta \frac{\langle k_{in} k_{out} \rangle}{\langle k \rangle}$ or $\delta < \beta \frac{\langle k_{in} k_{out} \rangle}{\langle k \rangle}$, regardless of the initial number of 'open' tasks. Here $\langle k_{in} k_{out} \rangle$ is the second joint moment of the in- and out-degree distributions (see Figure 8A and Appendix). We thus have a threshold phenomenon, similar to the random network case, which is further illustrated by the phase diagram shown in Figure 8B. The two phases in the diagram are now separated by the line $\delta = \beta \frac{\langle k_{in} k_{out} \rangle}{\langle k \rangle}$ (as opposed to the line $\delta = \beta \langle k \rangle$ in the random network case). This critical line can be interpreted as follows. Using the fact that the covariance between the in- and out-degrees of a node is $\text{cov}(k_{in}, k_{out}) = \langle k_{in} k_{out} \rangle - \langle k \rangle^2$, the critical line can also be written as:

$$\delta = \beta \frac{\langle k_{in} k_{out} \rangle}{\langle k \rangle} = \beta \left( \langle k \rangle + \frac{\text{cov}(k_{in}, k_{out})}{\langle k \rangle} \right) = \beta \langle k \rangle + \text{"correlations"} \qquad (2)$$

We thus see that the critical line is shifted by the amount of correlation (essentially related to covariance) between $k_{in}$ and $k_{out}$ in the network—a beautiful example of a network effect. If $k_{in}$ and $k_{out}$ are positively correlated (i.e. the network is assortative), the critical line is shifted upward relative to the critical line corresponding to an uncorrelated random network, $\delta = \beta \langle k \rangle$. This upward shift has a negative effect on projects; it shrinks the region corresponding to a convergent project (see Figure 8B), thereby reducing the number of available degrees of freedom, and increasing the likelihood that the project spirals out of control. To illustrate, consider the convergent project corresponding to point 1 in Figure 8B. Increasing the coupling between tasks (perhaps due to product redesign) even slightly may lead to a project that spirals out of control (point 2). We note that for uncorrelated random networks $\langle k_{in} k_{out} \rangle = \langle k_{in} \rangle \langle k_{out} \rangle = \langle k \rangle^2$, or equivalently $\text{cov}(k_{in}, k_{out}) = 0$. Plugging into equation 2 gives the critical threshold for Erdős-Rényi random networks $\delta = \beta \langle k \rangle$, as already derived above.

The critical line in Figure 8B can also be interpreted in the following way:

$$\delta = \beta \frac{\langle k_{in} k_{out} \rangle}{\langle k \rangle} = \beta \frac{\langle k_{in} k_{out} \rangle}{\langle k \rangle^2} \langle k \rangle = \beta' \langle k \rangle \qquad (3)$$

where $\beta' = \beta \frac{\langle k_{in} k_{out} \rangle}{\langle k \rangle^2}$ is the 'effective coupling.' Thus, we see that a correlated network with average degree $\langle k \rangle$, recovery coefficient $\delta$, and coupling strength $\beta$ has the same dynamical effect as a random network with the same average degree $\langle k \rangle$ and recovery coefficient $\delta$, but with an effective coupling $\beta'$. Moreover, if $k_{in}$ and $k_{out}$ are positively correlated (that is the network is assortative), $\langle k_{in} k_{out} \rangle > \langle k \rangle^2$ and thus $\beta' > \beta$. In other words, the faster propagation of errors resulting from positive correlations in the network is essentially equivalent (dynamically speaking) to an increase in the effective level of coupling and dependencies between nodes in a related random network.

The above analysis provides an explanation for the empirical results reported in Figure 4. There, it was shown that complex design networks tend to be uncorrelated or disassortative in nature; that is, complex design networks exhibit no (or even negative) correlations in their degree connectivity patterns. In light of our model, a negative (or no) correlation between $k_{in}$ and $k_{out}$ has the effect of shifting the critical line *downward* relative to the critical line corresponding to an uncorrelated random network. This will increase the number of available degrees of freedom, and decrease the likelihood that the project spirals out of control.

In summary, we presented a model of error dynamics and change propagation in complex design networks, and most importantly have demonstrated the deep relationship between the structure of networks and the resulting dynamics. We next apply the model to the study of two key properties of complex design networks: robustness and sensitivity.

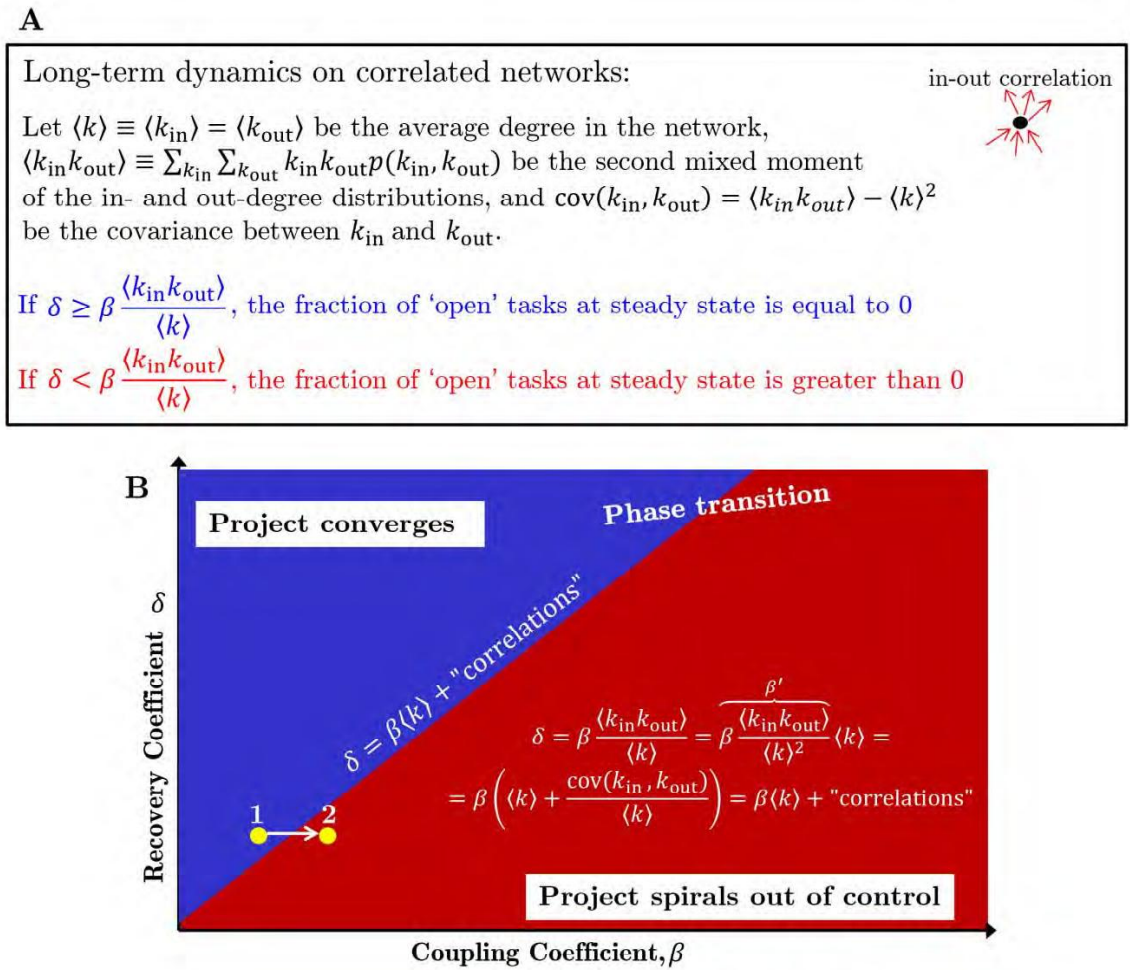

**Figure 8. The dynamics of error propagation on correlated networks.** Here we assume that the correlations between neighboring nodes are very small, and the only relevant correlation is the one between the in-degree and out-degree of individual nodes. (**A**) The long-term dynamics of the network is determined by a threshold that depends on the coupling coefficient, recovery coefficient, and the in-out correlation in the network. (**B**) The two phases in the diagram are separated by a critical line, which is shifted upward—by the amount of positive correlation in the network—relative to the critical line $\delta = \beta \langle k \rangle$ corresponding to an uncorrelated Erdős-Rényi random network.

# 5. Robustness and Leverage of Complex Design Networks

In this section, we discuss the functional role of the right-skewness and 'wild' variability characteristics of the connectivity distributions observed in complex design networks (see Figure 3). The first functional property—robustness—is the ability of a network to maintain its performance despite extreme, often unanticipated, events that affect the individual nodes in the network. The second functional property—leverage—is the ability to remarkably improve the performance of the network by preferentially allocating engineering resources to certain parts of the network. This is often achieved by prioritizing the efforts towards the highly connected nodes in the network (hubs). We demonstrate these two functional properties by simulating the dynamic network model presented in Section 5 on real-world design networks.

## 5.1 Robustness and Vulnerability

We illustrate the concept of robustness in the context of product development networks [14]. We measure the performance of the network in terms of the time it takes for the project to converge to the globally resolved state, where the fraction of 'open' tasks becomes zero (assuming the conditions for convergence are satisfied, see Section 5 and Figure 8).

We start with a 'normally running' project network where $\beta_i = \beta$ and $\delta_i = \delta$ for all nodes. To emulate extreme events that could occur over time, we select a fraction of nodes in the network and impair their characteristic parameters so that $\beta_i^{\text{new}} > \beta$ or $\delta_i^{\text{new}} < \delta$. The former modification reflects, for example, changes in product design that lead to increased coupling between tasks in the project network. The latter case reflects, for example, changes in project resources that lead to development tasks that require longer autonomous development times.

We consider several rules of selecting the nodes that will be impaired in the network. The first rule is to select the nodes randomly, regardless of their structural position in the network. We can also prioritize the nodes according to some rule that takes into account their structural position in the network. Here we consider four priority rules: **Fan-in**, select tasks in decreasing order of their in-degrees; **Fan-out,** select tasks in decreasing order of their out-degrees; **Sum,** select tasks in decreasing order of their total degree (sum of in-and out-degrees); **Product**, select tasks in decreasing order of the product of their in-degree and out-degrees. Starting with a 'normally running' project network and impairing a percentage of nodes in the network will clearly impair the performance of the network (in our case, prolonging the duration of the project). The question that we ask here is whether or not the above rules affect the network performance to the same extent. Remarkably, as shown in Figure 9, we find that the network performance is extremely robust if nodes are impaired in a random order; that is, the duration of the project goes up very slowly as increasingly more nodes of the network are impaired. However, a completely different behavior is observed if nodes are impaired according to the above priority rules. As increasingly more nodes of the network are impaired, the duration of the project is increased rapidly and dramatically, becoming about twice longer as its original value even if only 6% of the tasks are impaired (see Figure 9). Thus, the network performance becomes highly sensitive to changes targeted at highly connected nodes. These findings apply to all of the design networks included in Table 2. We can sum up these observations as follows. The dynamics of design networks is *ultra-robust* and *error tolerant* when negative design changes occur at randomly selected nodes; yet *highly vulnerable* and *fragile* when unwanted changes are targeted at highly central nodes.

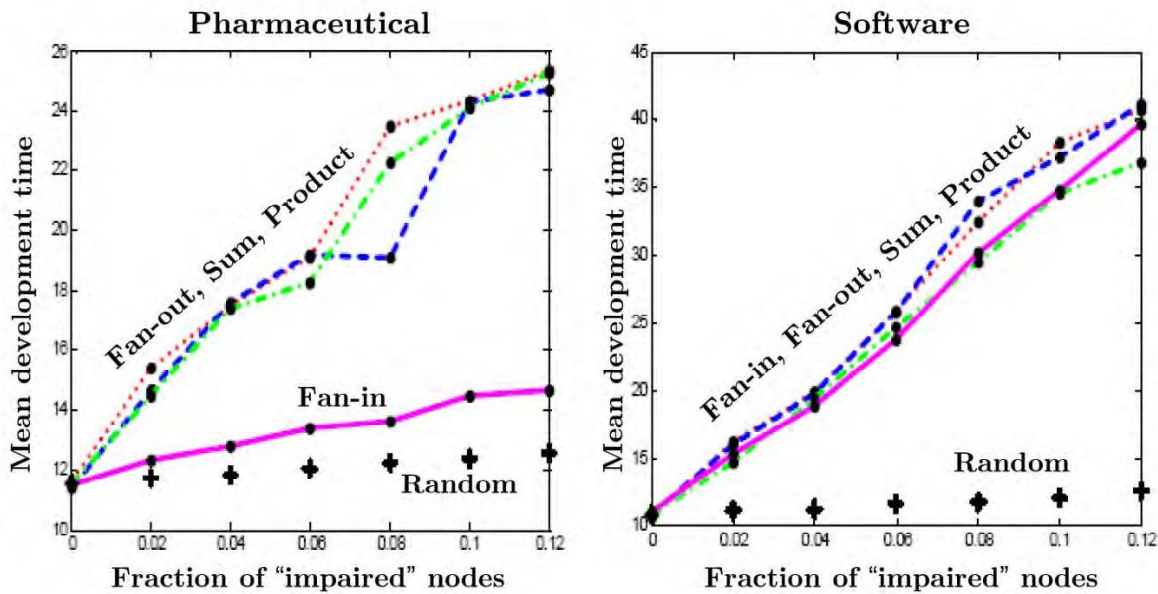

**Figure 9. Robustness and vulnerability of complex design networks.** We compare the five priority rules: Random (+), Fan-in (magenta solid line), Fan-out (green dash-dot line), Sum (blue dashed line), and Product (red dotted line). The figure presents the network performance (project duration) versus the fraction of impaired tasks in the network for which their coupling coefficients are modified. For the non-random priority rules, each data point is the average of 1000 realizations. For the Random rule, each point is the average of 30 different task selections, performed for 100 independent runs. The model parameters, before and after the change, are as follows: **Software**. $\delta = 0.75$, $\beta = 0.05$, $\beta^{new} = 0.1$; **Pharmaceutical**. $\delta = 0.75$, $\beta = 0.05$, $\beta^{new} = 0.1$.

## 5.2 Leverage and Control

The robustness characteristic of complex design networks deals mostly with unexpected adverse changes that could occur in the network. The dual concept of leverage deals with deliberate network changes that aim at improving and controlling the performance of the design network. More specifically, the sensitivity of the network to changes directed at highly connected nodes can be utilized by designers to influence the performance of the network [14]. The structure of our analysis is similar to that of the previous section; the only difference is that now we select a fraction of nodes in the network and improve (rather than impair) their characteristic parameters so that now $\beta_i^{\text{new}} < \beta$ or $\delta_i^{\text{new}} > \delta$. The former modification reflects, for example, changes in product design that lead to modular architectures and reduced dependencies between tasks in the project network. The latter case reflects, for example, allocation of additional project resources that lead to development tasks that require shorter development times.

The results are presented in Figure 10. When nodes in the network are selected in a random order, we find that the performance of the project is improved very slowly; that is, the duration of the project goes down gradually as more nodes of the network are increasingly modified and improved. However, a drastically different behavior is observed when tasks are selected based on a preferential policy that takes into account their connectivity in the network (e.g. using the Fan-in, Fan-out, Sum, or Product rules). In this case, as increasingly more nodes of the network are improved, the duration of the project is decreased rapidly and dramatically, becoming about twice shorter as its original value even if only 6% of the tasks are improved (see Figure 10). This remarkable behavior is observed for all of the design networks included in Table 2.

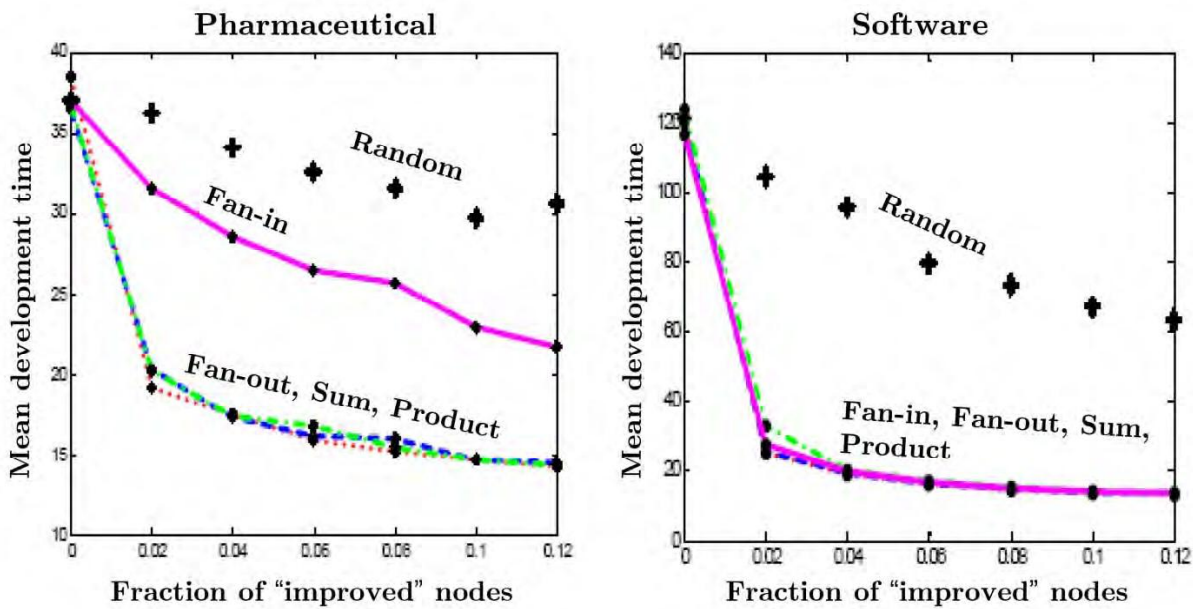

**Figure 10. Leverage and control of complex design networks.** We compare the five priority rules: Random (+), Fan-in (magenta solid line), Fan-out (green dash-dot line), Sum (blue dashed line), and Product (red dotted line). The figure presents the network performance (project duration) versus the fraction of improved tasks in the network for which their coupling coefficients are modified. For the non-random priority rules, each data point is the average of 1000 realizations. For the Random rule, each point is the average of 30 different task selections, performed for 100 independent runs. The model parameters, before and after the change, are as follows: **Software.** $\delta = 0.75$, $\beta = 0.1$, $\beta^{\text{new}} = 0.05$; **Pharmaceutical.** $\delta = 0.75$, $\beta = 0.1$, $\beta^{\text{new}} = 0.05$.

In sum, the heavy-tailed degree distributions and the characteristic feature of 'hubs' (highly connected nodes) offer a strategy for exploiting complex design networks—a remarkable improvement in the performance of the system can be achieved by focusing engineering and management efforts on central nodes in the network. Simultaneously, the long right-tail of the degree distributions also leads to robustness under the circumstances that unanticipated and negative changes affect nodes in a random fashion (see Section 5.1). On the other hand, the long right-tail also makes the network more fragile and vulnerable to unanticipated negative changes that occur at highly connected nodes (see Section 5.1); a condition that could lead to failure and spiraling out of control. This 'no free lunch' principle lies at the heart of complex design networks.

# 6. Summary

Large-scale engineering design often involves hundreds or thousands of designers that self-organize to develop, tweak and tinker architectural designs, which are locally optimized to be integrated in the larger system. The remarkable thing is that this tinkering process leads to large-scale universal patterns and system properties that were not written in the initial specification sheet or anticipated from the outset. Here we described a wide variety of large-scale design networks—including open source software, electronic circuits, product development, power grids, and the internet. These systems share common structural properties of networks such as sparseness, heavy-tailed degree distributions, high clustering coefficients, short average path lengths between any two nodes, and negative (or no) correlations in their degree connectivity patterns (disassortative mixing by degree).

We presented and analyzed a model for the dynamics of errors, rework, or change propagation in complex design networks. The model is based on the idea that non-trivial, large-scale behavior can be produced by simple processes involving interactions between the nodes in the network. The key result of our model is that the network structure provides direct information about the characteristics of error and rework dynamics. For example, in the context of product design and development, the dynamics is characterized by a phase transition from convergence to the globally resolved state, where the fraction of 'open' tasks becomes zero, to the state where the project spirals out of control with persistent 'open' tasks in the network. The threshold separating the two phases was found to be closely related to the extent of degree correlations in the network; in particular, positively correlated networks tend to impede the convergence of product development processes. The heavy-tailed degree distributions and the existence of hubs affect the functionality of design networks in intricate ways. First, the dynamic behavior of complex design networks is highly robust to uncontrolled changes occurring at random nodes; yet vulnerable to changes that are targeted at central nodes. At the same time, changes that are directed at highly connected nodes can significantly boost the performance and efficiency of the network.

The emerging discipline of complex networks research offers a new and potentially powerful perspective on managing large-scale engineering design systems. By mapping the information flows underlying large-scale systems, supported by network visualization and modelling tools, engineers could gain better understanding on the intricate relationship between structure and dynamics. We anticipate that the empirical, theoretical, and practical insights gained by modeling large-scale engineering design systems as self-organizing complex networks will turn out to be highly relevant to the science of design.

## Appendix. Measuring Complex Networks

Complex networks can be defined formally in terms of a graph $G = (V, E)$, which is a set of nodes $V = \{1, 2, \cdots, N\}$, and a set of lines $E = \{e_1, e_2, \cdots e_L\}$ between pairs of nodes. If the line between two nodes is non-directional, then the network is called undirected; otherwise, the network is called directed. A network is usually represented by a diagram, where nodes are drawn as points, undirected lines are drawn as edges and directed lines as arcs connecting the corresponding two nodes. Several properties have been used to characterize 'real-world' complex networks:

**Density**. The density $D$ of a network is defined as the ratio between the number of edges (arcs) $L$ to the number of possible edges (arcs) in the network:

$$D = \frac{2L}{N(N-1)} \text{ (undirected networks)} \qquad D = \frac{L}{N(N-1)} \text{ (directed networks)} \qquad (A1)$$

**Characteristic Path Length**. The average distance (geodesic) $d(i,j)$ between two nodes $i$ and $j$ is defined as the number of edges along the shortest path connecting them. The characteristic path length $d$ is the average distance between any two vertices in the network:

$$d = \frac{1}{N(N-1)} \sum_{i \neq j} d(i,j) \qquad (A2)$$

**Clustering Coefficient**. The clustering coefficient measures the tendency of nodes to be locally interconnected or to cluster in dense modules. Let node $i$ be connected to $k_i$ neighbors. The total number of edges between these neighbors is at most $k_i(k_i - 1)/2$. If the actual number of edges between these $k_i$ neighbors is $n_i$, then the clustering coefficient $C_i$ of a node $i$ is the ratio

$$C_i = \frac{n_i}{k_i(k_i-1)/2} \qquad (A3)$$

The clustering coefficient of the entire network, which is a global measure of the network's potential modularity, is the average over all nodes,

$$C = \frac{\sum_{i=1}^{N} C_i}{N} \tag{A4}$$

**Degree Centrality**. The degree of a vertex, denoted by $k_i$, is the number of nodes adjacent to it. The mean node degree (often called the first moment) is the average degree of the nodes in the network,

$$\langle k \rangle = \frac{\sum_{i=1}^{N} k_i}{N} = \frac{2L}{N} \tag{A5}$$

If the network is directed, a distinction is made between the in-degree of a node and its out-degree. The in-degree of a node, $k_{\text{in}}(i)$, is the number of nodes that are adjacent to $i$. The out-degree of a node, $k_{\text{out}}(i)$, is the number of nodes adjacent from $i$. For directed networks, $\langle k_{\text{in}} \rangle = \langle k_{\text{out}} \rangle = \langle k \rangle$. Other node centrality indices were established, including closeness centrality, betweenness centrality, and eigenvector centrality [12].

**Degree Distribution**. The node degree distribution $p(k)$ is the probability that a node has $k$ edges. The corresponding in- or out-degree distributions for directed networks are denoted by $p_{\text{in}}(k)$ and $p_{\text{out}}(k)$, respectively.

**Connected Components**. A Weakly (Strongly) Connected Component is a set of nodes in which there exists an undirected (directed) path from any node to any other. The single connected component which contains most of the nodes in the network (and thus many loops) is referred to as the Giant Component. For a certain class of networks in which degrees of nearest-neighbor nodes are not correlated, the critical thresholds for the existence of a giant component is found by the following criteria:

$$\frac{\langle k^2 \rangle}{\langle k \rangle} \geq 2 \text{ (undirected networks)} \quad \frac{\langle k_{\text{in}} k_{\text{out}} \rangle}{\langle k \rangle} \geq 1 \text{ (directed networks)} \tag{A6}$$

where $\langle k^2 \rangle$ and $\langle k_{\text{in}} k_{\text{out}} \rangle$ are the second moment and joint moment of the degree distributions, respectively. We notice that, for undirected networks, higher variability of the degree distribution leads to a giant component. For directed networks, a higher correlation between the in- and out-degrees of nodes leads to a giant component, which in turn leads to complicated network dynamics (see main text).